\documentclass[9pt,twocolumn,twoside]{pnas-new}
\templatetype{pnasresearcharticle} % Choose template 
\setboolean{displaywatermark}{false}

\title{The vortex gas scaling regime of baroclinic turbulence}

\author[a,1]{Basile Gallet}
\author[b]{Raffaele Ferrari} 

\affil[a]{Service de Physique de l'Etat Condens\'e, CEA Saclay, CNRS UMR 3680, Universit\'e Paris-Saclay, 91191 Gif-sur-Yvette, France.}
\affil[b]{Massachusetts Institute of Technology, USA}

\leadauthor{B. Gallet} 

\significancestatement{Developing a theory of climate requires an accurate parameterization of the transport induced by turbulent eddies. A major source of turbulence in the mid-latitude planetary atmospheres and oceans is the baroclinic instability of the large-scale flows. We present a scaling theory that quantitatively predicts the local heat flux, eddy kinetic energy and mixing length of baroclinic turbulence as a function of the large-scale flow characteristics and bottom friction. The theory is then used as a quantitative parameterization in the case of meridionally dependent forcing, in the fully turbulent regime. Beyond its relevance for climate theories, our work is an intriguing example of a successful closure for a fully turbulent flow.}

\authordeclaration{The authors declare no conflict of interest.}
\equalauthors{\textsuperscript{1} All the authors contributed equally to this work.}
\correspondingauthor{\textsuperscript{2} To whom correspondence should be addressed. E-mail: basile.gallet@cea.fr}

\keywords{Oceanography $|$ Atmospheric dynamics $|$ Turbulence} 

\newcommand{\la}{\left<}
\newcommand{\ra}{\right>}
\newcommand{\liv}{\ell_\text{iv}}

\newcommand{\cor}[1]{{#1}}

\begin{abstract}

The mean state of the atmosphere and ocean is set through a balance between external forcing -- radiative processes in the atmosphere and air-sea fluxes of momentum, heat and freshwater in the ocean -- and the emergent turbulence which transfers energy to dissipative structures, primarily through friction in bottom boundary layers. The external forcing maintains lateral temperature gradients, which on a rotating planet give rise to flows along the temperature contours: jets in the atmosphere and currents in the ocean. These large-scale flows spontaneously develop turbulent eddies through the baroclinic instability. A critical step in the development of a theory of climate is to properly include the resulting eddy-induced turbulent transport of properties like heat, moisture, and carbon. In the early linear stages, baroclinic instability generates flow structures at the Rossby deformation radius, a length scale of order 1000 km in the atmosphere and 100 km in the ocean, smaller than the planetary scale and much smaller than the typical extent of ocean basins respectively. There is therefore a separation of scales, arguably more in the ocean than in the atmosphere, between the large-scale temperature gradient and the smaller eddies that advect it randomly, inducing effective diffusion. Numerical solutions of the two-layer quasi-geostrophic model, the standard model for studies of eddy motions in the atmosphere and ocean, show that such scale separation remains in the strongly nonlinear turbulent regime, provided there is sufficient bottom drag.

\cor{We compute the scaling-laws governing the eddy-driven transport associated with baroclinic turbulence. First, we provide a theoretical underpinning for empirical scaling-laws reported in previous studies, for different formulations of the bottom drag law. Secondly, these scaling-laws are shown to provide an important first step toward an accurate local closure to predict the impact of baroclinic turbulence in setting the large-scale temperature profiles in the atmosphere and ocean.}
\end{abstract}

\dates{This manuscript was compiled on \today}
\doi{\url{www.pnas.org/cgi/doi/10.1073/pnas.XXXXXXXXXX}}

\begin{document}

\maketitle
\thispagestyle{firststyle}
\ifthenelse{\boolean{shortarticle}}{\ifthenelse{\boolean{singlecolumn}}{\abscontentformatted}{\abscontent}}{}

\begin{figure*}[t]
\centering
\includegraphics[width=18 cm]{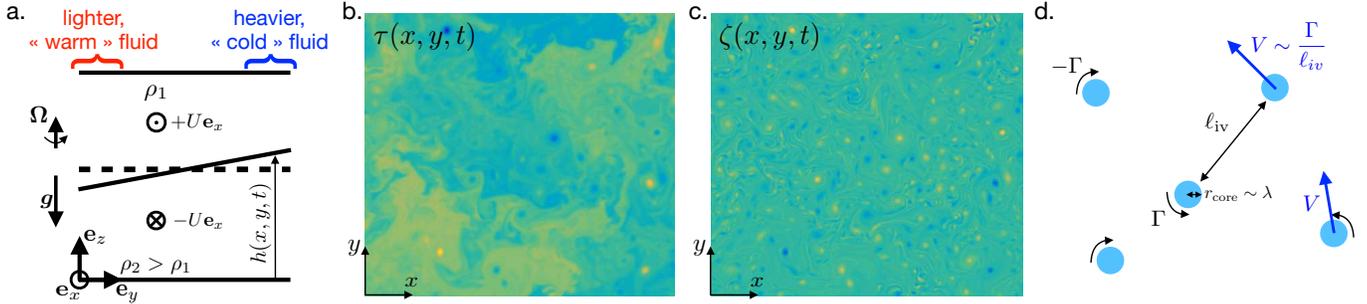}
\caption{panel a: base state of the 2LQG system with imposed vertical shear. The interface is tilted in the $y$ direction as a consequence of thermal wind balance. The baroclinic streamfunction is proportional to $-h$, where $h(x,y,t)$ is the local  displacement of the interface. For this reason, the baroclinic streamfunction is often referred to as the ``temperature'' field. Snapshots of the departure of the baroclinic streamfunction from the base state ($\tau$, panel b) and of the barotropic vorticity ($\zeta$, panel c) from a numerical simulation in the low-friction regime (arbitrary units, low values in dark blue and large values in bright yellow). We model the barotropic flow as a gas of vortices (panel d) of circulation $\pm \Gamma$ and radius $r_\text{core}\sim \lambda$. The vortex cores move as a result of their mutual interaction, with a typical velocity $V\sim \Gamma/\liv$, where $\liv$ is the typical inter-vortex distance.}\label{fig:gas}
\end{figure*}

\dropcap{O}ceanic and atmospheric flows are subject to the combined effects of strong density stratification and rapid planetary rotation. On the one hand, these two ingredients add complexity to the dynamics, making the flow strongly anisotropic and inducing waves that modify the characteristics of the turbulent eddies. On the other hand, they permit the derivation of reduced sets of equations that capture the large-scale behavior of the flow: this is the realm of quasi-geostrophy (QG). The outcome of this approach is a model that couples two-dimensional layers of fluid of different density. QG filters out fast-wave dynamics, relaxing the necessity to resolve the fastest time scales of the original system. A QG model with only two fluid layers is simple enough for fast and extensive numerical studies, and yet it retains the key phenomenon arising from the combination of stable stratification and rapid rotation~\cite{Flierl}: baroclinic instability, with its ability to induce small-scale turbulent eddies from a large-scale vertically sheared flow. The two-layer quasi-geostrophic model (2LQG) offers a testbed to derive and validate closure models for the ``baroclinic turbulence'' that results from this instability.% while being a surprisingly good model for large-scale oceanic and atmospheric flows, despite its simplicity.

In the simplest picture of 2LQG, a layer of light fluid sits on top of a layer of heavy fluid, as sketched in Fig.~\ref{fig:gas}a, \cor{in a frame rotating at a spatially uniform rate $\Omega=f/2$ around the vertical axis. Such a uniform Coriolis parameter $f$ is a strong simplification as compared to real atmospheres and oceans, where the $\beta$-effect associated with latitudinal variations in $f$ can trigger the emergence of zonal jets. Nevertheless, $\beta$ vanishes at the poles of a planet, and it seems than any global parameterization of baroclinic turbulence needs to correctly handle the limiting case $\beta=0$, which we address in the present study. The 2LQG model applies to motions evolving on timescales long compared to the planetary rotation -- the small-Rossby-number limit -- and on horizontal scales larger than the equal depths of the two layers; see Ref. \cite{Salmonbook,Vallisbook} for more details on the derivation of QG. At leading order in Rossby number the vertical momentum equation reduces to hydrostatic balance\footnote{Hydrostatic balance is the balance between the upward-directed pressure gradient force and the downward-directed force of gravity.}, while the horizontal flow is in geostrophic balance\footnote{Geostrophic balance is the balance between the Coriolis force and lateral pressure gradient forces.}. These two balances imply that both the flow field and the local thickness of each layer can be expressed in terms of the corresponding streamfunctions, $\psi_1(x,y,t)$ in the upper layer and $\psi_2(x,y,t)$ in the lower layer. At the next order in Rossby number, the vertical vorticity equation yields the evolution equations for $\psi_1(x,y,t)$ and $\psi_2(x,y,t)$:}
\begin{eqnarray}
\partial_t q_1 + J(\psi_1,q_1) & = &  -\nu \Delta^4 q_1 \, , \label{eqq1}\\
\partial_t q_2 + J(\psi_2,q_2) & = & -\nu \Delta^4 q_2 + \text{drag} \label{eqq2} \, ,
\end{eqnarray}
\cor{where the subscripts $1$ and $2$ refer again to the upper and lower layers, and the Jacobian is  $J(f,g)=\partial_x f \partial_y g - \partial_x g \partial_y f$. The potential vorticities $q_1(x,y,t)$ and $q_2(x,y,t)$ are related to the streamfunctions through:}
\begin{eqnarray}
q_1 & = & \boldsymbol{\nabla}^2 \psi_1 + \frac{1}{2\lambda^2}(\psi_2-\psi_1) \, ,\\
q_2 & = & \boldsymbol{\nabla}^2 \psi_2 + \frac{1}{2\lambda^2}(\psi_1-\psi_2) \, ,
\end{eqnarray}
\cor{where $\lambda$ denotes the Rossby deformation radius\footnote{The Rossby radius of deformation $\lambda$ is the length scale at which rotational effects become as important as buoyancy or gravity wave effects in the evolution of a flow.}. In our model, the drag term is confined to the lower-layer equation [\ref{eqq2}]. In the case of linear drag, $\text{drag} =-2 \kappa \boldsymbol{\nabla}^2 \psi_2 $, and in the case of quadratic drag, $\text{drag} =- \mu \left[ \partial_x(|\boldsymbol{\nabla} \psi_2| \partial_x \psi_2) + \partial_y(|\boldsymbol{\nabla} \psi_2| \partial_y \psi_2)  \right]$. Finally, equations [\ref{eqq1}] and [\ref{eqq2}] include hyperviscosity to dissipate filaments of potential vorticity (enstrophy) generated by eddy stirring at small scales.}

%The governing equations can be written in terms of the velocity streamfunctions\footnote{A streamfunction $\psi$ is a function whose isolines are everywhere tangent to the velocity field. The zonal (along $x$) and meridional (along $y$) velocities are given by $u \equiv -\partial_y \psi$ and $v\equiv\partial_x \psi$.} of both layers, but 
\cor{A more insightful representation arises from the sum and difference of equations [\ref{eqq1}] and [\ref{eqq2}]:} one obtains an evolution equation for the barotropic streamfunction -- half the sum of the streamfunctions of both layers -- which characterizes the vertically invariant part of the flow, and an evolution equation for the baroclinic streamfunction -- half the difference between the two streamfunctions -- which characterizes the vertically dependent flow. Because in QG the streamfunction is directly proportional to the thickness of the fluid layer, the baroclinic streamfunction is also a measure of the height of the interface between the two layers. A region with large baroclinic streamfunction corresponds to a locally deeper upper layer: there is more light fluid at this location, and we may thus say that on vertical average the fluid is warmer. Similarly, a region of low baroclinic streamfunction corresponds to a locally shallower upper layer, with more heavy fluid: this is a cold region. Thus, the baroclinic streamfunction is often denoted as $\tau$ and referred to as the local ``temperature'' of the fluid.

The 2LQG model can be used to study the equilibration of baroclinic instability arising from a prescribed horizontally uniform vertical shear, which represents the large-scale flows maintained by external forcing in the ocean and atmosphere. Denoting the vertical axis as $z$ and the zonal and meridional directions as $x$ and $y$, the prescribed flow in the upper and lower layers consists respectively in zonal motion $+U {\bf e}_x$ and $-U {\bf e}_x$. This flow is in thermal wind balance\footnote{A flow is in thermal wind balance if frictional forces and accelerations are weak, except for the Coriolis acceleration associated with Earth's rotation.} with a prescribed uniform meridional temperature gradient $-U$, i.e. there is a sloping interface between the heavy and light fluid layers, see Fig.~\ref{fig:gas}a. This tilt provides an energy reservoir, the Available Potential Energy~\cite[APE,][]{Lorenz}, that is released by baroclinic instability acting to flatten the density interface. We denote respectively as $\psi(x,y,t)$ and $\tau(x,y,t)$ the perturbations of barotropic and baroclinic streamfunctions around this base state, and consider their evolution equations inside a large (horizontal) domain with periodic boundary conditions in both $x$ and $y$:
\begin{eqnarray}
& & \partial_t  (\boldsymbol{\nabla}^2 \psi) + J(\psi,\boldsymbol{\nabla}^2 \psi) + J(\tau,\boldsymbol{\nabla}^2 \tau) + U \partial_x (\boldsymbol{\nabla}^2 \tau)  \label{eqpsi} \\
\nonumber  & & =  -\nu \boldsymbol{\nabla}^{10} \psi  + \text{drag}/2 \, ,   \\
& & \partial_t  [\boldsymbol{\nabla}^2 \tau - \lambda^{-2} \tau] + J(\psi,\boldsymbol{\nabla}^2 \tau - \lambda^{-2} \tau) + J(\tau,\boldsymbol{\nabla}^2 \psi)  \label{eqtau} \\
\nonumber  & & + U \partial_x [ \boldsymbol{\nabla}^2 \psi + \lambda^{-2} \psi ] = -\nu \boldsymbol{\nabla}^{8} [ \boldsymbol{\nabla}^2 \tau - \lambda^{-2} \tau]  - \text{drag}/2 \, .
\end{eqnarray}
The system releases APE by developing eddy motion through baroclinic instability, and the goal is to characterize the statistically steady turbulent state that ensues: how energetic is the barotropic flow? How strong are the local temperature fluctuations? And, most importantly, what is the eddy-induced meridional heat-flux? The latter quantity is a key missing ingredient required to formulate a theory of the mean state of the atmosphere and ocean as a function of external forcing parameters~\cite{Held}.

Traditionally, these questions have been addressed using descriptions of the flow in spectral space, focusing on the cascading behavior of the various invariants~\cite{Salmon}. In contrast with this approach, Thompson \& Young~\cite[][TY in the following]{Thompson} describe the system in physical space and argue that the barotropic flow evolves towards a gas of isolated vortices. In spite of this intuition, TY cannot conclude on the scaling behavior of the quantities mentioned above and resort to empirical fits instead. Focusing on the case of linear drag, they conclude that the temperature fluctuations and meridional heat flux are extremely sensitive to the drag coefficient: they scale exponentially in inverse drag coefficient. This scaling dependence was recently shown by Chang \& Held~\cite[][CH in the following]{Chang} to change quite drastically if linear drag is replaced by quadratic drag: the exponential dependence becomes a power-law dependence on the drag coefficient. However, CH acknowledge the failure of standard cascade arguments to predict the exponents of these power laws, and they resort to curve fitting as well.

In this Letter, we supplement the vortex gas approach of TY with statistical arguments from point vortex dynamics to obtain a predictive scaling theory for the eddy kinetic energy, the temperature fluctuations and the meridional heat flux of baroclinic turbulence. The resulting scaling theory captures both the exponential dependence of these quantities on the inverse linear drag coefficient, and their power-law dependence on the quadratic drag coefficient. Our predictions are thus in quantitative agreement with the scaling-laws diagnosed by both TY and CH. Following Pavan \& Held \cite{Pavan} and CH, we finally show how these scaling-laws can be used as a quantitative turbulent closure to make analytical predictions in situations where the system is subject to inhomogeneous forcing at large scale.

\begin{figure*}[h!]
%\vspace{-3 cm}
    \centerline{\includegraphics[width=18 cm]{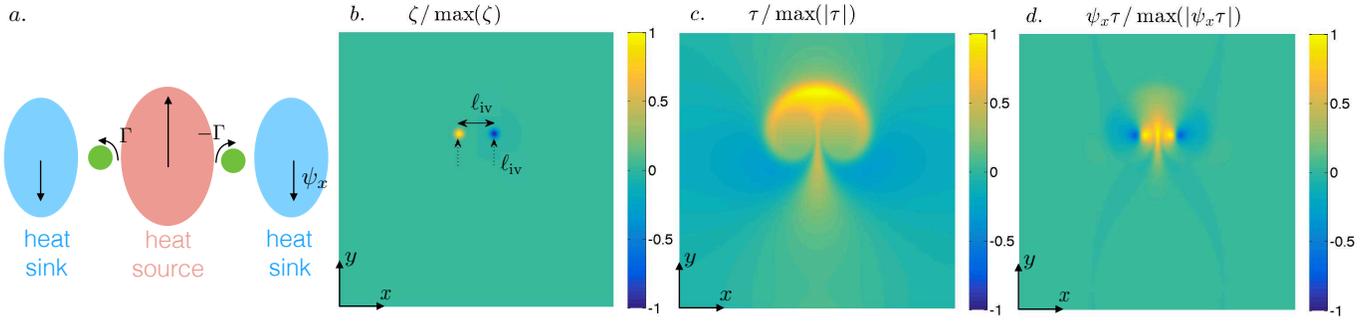} }
   \caption{\textbf{Heat transport by a barotropic vortex dipole.} Panel a is a schematic representation of the heat sources and sinks induced by the dipolar velocity field. Panels b, c and d show respectively the barotropic vorticity, temperature field, and local meridional heat flux, at the end time of a numerical solution of [\ref{tauLS}] where the dipole travels over a distance $\liv$ in the meridional direction $y$. \label{figdipole}}
\end{figure*}

\section*{The QG vortex gas}

Denoting as $\la \cdot \ra$ a spatial and time average and as $\psi_x=\partial_x \psi$ the meridional barotropic velocity, our goal is to determine the meridional heat flux $\la \psi_x \tau \ra$, or equivalently the diffusivity $D=\la \psi_x \tau \ra/U$ that connects this heat flux to minus the background temperature gradient $U$. A related quantity of interest is the mixing-length $\ell=\sqrt{\la \tau^2 \ra}/U$. This is the typical distance travelled by a fluid element carrying its background temperature, before it is mixed with the environment and relaxes to the local background temperature. It follows that the typical temperature fluctuations around the background gradient are of the order of $U \ell$. We seek the dependence of $D$ and $\ell$ on the various external parameters of the system. It was established by TY that, for a sufficiently large domain, the mixing-length saturates at a value much smaller than the domain size, and independent of it. The consequence is that the size of the domain is irrelevant for large enough domains. The small-scale dissipation coefficient -- a hyperviscosity in most studies -- is also shown by TY to be irrelevant when low enough. %Roughly speaking, this is a consequence of the inverse cascade of the barotropic energy, together with the anomalous enstrophy dissipation at small scale, both of which do not involve the value of hyperviscosity, see [Salmon???]. 
The quantities $D$ and $\ell$ thus depend only on the dimensional parameters $U$, the Rossby deformation radius $\lambda$, and the bottom drag coefficient, denoted as $\kappa$ in the case of linear friction (with dimension of an inverse time) and as $\mu$ in the case of quadratic drag (with dimension of an inverse length)~\cite{Arbic,Chang}. %Additional details on these notations, as well as the implementation of drag in the equations, are provided in the Supplementary Information (SI). 
In dimensionless form, we thus seek the dependence of the dimensionless diffusivity $D_*=D/U \lambda$ and mixing-length $\ell_*=\ell/\lambda$ on the dimensionless drag $\kappa_*=\kappa \lambda /U$ or $\mu_*=\mu \lambda$.

We follow the key intuition of TY that the flow is better described in physical space than in spectral space. In Fig.~\ref{fig:gas}, we provide snapshots of the barotropic vorticity and baroclinic streamfunction from a direct numerical simulation in the low-drag regime (see the numerical methods in the Supplementary Information, SI): the barotropic flow consists of a ``gas'' of well-defined vortices, with a core radius substantially smaller than the inter-vortex distance $\ell_{iv}$. \cor{Vortex gas models were introduced to describe decaying two-dimensional (purely barotropic) turbulence. It was shown that the time evolution of the gross vortex statistics, such as the typical vortex radius and circulation, can be captured using simpler ``punctuated Hamiltonian'' models~\cite{Carnevale,WeissMcWilliams,Trizac}. The latter consist in integrating the Hamiltonian dynamics governing the interaction of localized compact vortices~\cite{Onsager}, interrupted by instantaneous merging events when two vortices come close enough to one another, with specific merging rules governing the strength and radius of the vortex resulting from the merger. These models were adapted to the forced-dissipative situation by Weiss~\cite{Weiss}, through injection of small vortices with a core radius comparable to the injection scale, and removal of the largest vortices above a cut-off vortex radius. The resulting model captures the statistically steady distribution of vortex core radius observed in direct numerical simulations of the 2D Navier-Stokes equations~\cite{Borue}: $P(r_\text{core}) \sim r_\text{core}^{-4}$ for $r_\text{core}$ above the injection scale. One immediate consequence of such a steeply decreasing distribution function is that the mean vortex core radius is comparable to the injection scale. Similarly, the average vortex circulation is dominated by the circulation of the injected vortices. In the following, we will thus infer the transport properties of the barotropic component of the two-layer model by focusing on an idealised vortex gas consisting of vortices with a single ``typical'' value of the vortex core radius $r_\text{core}$ comparable to the injection scale, and circulations $\pm \Gamma$, where $\Gamma$ is the typical magnitude of the vortex circulation. For baroclinic turbulence, both linear stability analysis \cite{Salmonbook,Vallisbook} and the multiple cascade picture \cite{Salmon} indicate that the barotropic flow receives energy at a scale comparable to the deformation radius $\lambda$. As discussed above, the typical vortex core radius is comparable to this injection scale, and we obtain $r_\text{core} \sim \lambda$. We stress the fact that such a small core radius is fully compatible with the phenomenology of the inverse energy cascade: inverse energy transfers result in the vortices being further appart, with little increase in core radius. The resulting velocity structures have a scale comparable to the large inter-vortex distance, even though the intense vortices visible in the vorticity field have a small core radius, comparable to the injection scale. Finally, it is worth noting that there is encouraging observational evidence both in the atmosphere and ocean that eddies have a core radius close to the scale at which they are generated through baroclinic instability~\cite{Schneider,Tulloch}. }

A schematic of the resulting idealized vortex gas is provided in Fig.~\ref{fig:gas}d: we represent the barotropic flow as a collection of vortices of circulation $\pm \Gamma$ and of core radius $r_\text{core}\sim \lambda$, and thus a velocity decaying as $\pm \Gamma /r$ outside the core~\cite{Charney63}. %The latter assumption is rather well satisfied in our numerical simulations, provided the hyperviscosity is sufficiently low. 
The vortices move as a result of their mutual interactions, with a typical velocity $V  \sim \Gamma / \ell_{iv}$. Through this vortex gas picture, we have introduced two additional parameters, $\liv$ and $V$ -- or, alternatively, $\Gamma=\liv \, V$ -- for a total of five parameters: $D$, $\ell$, $\ell_\text{iv}$, $V$, and a drag coefficient ($\kappa$ or $\mu$). We thus need four relations between these five quantities to produce a fully closed scaling theory.

%However, our description of the vortex gas departs from that of TY in two aspects: first, we assume that the vortex core radius simply scales with the Rossby deformation radius $\lambda$. This seems rather well satisfied in our numerical simulations, provided the hyperviscosity is sufficiently low. Second, we do not assume that the heat is transported within the vortex cores. Indeed, the  snapshot \ref{fig:gas}b makes it clear that $\tau$ is significantly nonzero outside the vortex cores. Because the vortex cores represent a very small areal fraction of the fluid domain, they have a negligible contribution to the overall meridional heat flux.

The first of these relations is the energy budget: the meridional heat flux corresponds to a rate of release of APE, $U  \la \psi_x \tau \ra / \lambda^2 = D U^2 / \lambda^2$, which is balanced by frictional dissipation of kinetic energy in statistically steady state. The contribution from the barotropic flow dominates this frictional dissipation in the low-drag asymptotic limit, and the energy power integral reads (see e.g. TY, and the SI):
\begin{eqnarray}
 \frac{D U^2}{\lambda^2} =  \left\{ \begin{matrix}
 \kappa \la {\bf u}^2 \ra \ & \text{for linear drag     } \, , \\
 \frac{\mu}{2} \la |{\bf u}|^3 \ra \ & \text{for quadratic drag} \, ,
\end{matrix} \right. \label{Ebudget}
\end{eqnarray}
%\textcolor{red}{[Facteur $1/2$ dans le cas quadratique?]}
where ${\bf u}=-\boldsymbol{\nabla}\times (\psi \, {\bf e}_z)$ denotes the barotropic velocity field. Our approach departs from both TY and CH in the way we evaluate the velocity statistics that appear on the right-hand side: we argue that a key aspect of vortex gas dynamics is that the various velocity moments scale differently, and cannot be estimated simply as $V$ above. Indeed, consider a single vortex within the vortex gas. It occupies a region of the fluid domain of typical extent $\liv$. The vorticity is contained inside a core of radius $r_\text{core} \sim \lambda \ll \liv$, and the barotropic velocity ${\bf u}$ has a magnitude $\Gamma/2 \pi r$ outside the vortex core, where $r$ is the distance to the vortex center. The velocity variance is thus:
\begin{equation}
\la {\bf u}^2 \ra = \frac{1}{\pi \liv^2} \int_{r_\text{core}}^{\liv} \frac{\Gamma^2}{4 \pi^2 r^2} \, 2 \pi r {\rm d} r \sim V^2 \log \left( \frac{\liv}{\lambda} \right) \, . \label{velvariance}
\end{equation}
This estimate for $\la {\bf u}^2 \ra$ exceeds that of TY by a logarithmic correction that captures the fact that the velocity is strongest close to the core of the vortex. This correction will turn out to be crucial to obtain the right scaling behaviors for $D_*$ and $\ell_*$. In a similar fashion, we estimate the third-order moment of the barotropic velocity field as:
\begin{equation}
\la |{\bf u}|^3 \ra = \frac{1}{\pi \liv^2} \int_{r_\text{core}}^{\liv} \frac{\Gamma^3}{8 \pi^3 r^3} \, 2 \pi r {\rm d} r \sim V^3 \, \frac{\liv}{\lambda}  \, , \label{vel3}
\end{equation}
where we have used the fact that $r_\text{core}\sim\lambda \ll \liv$. Again, this estimate exceeds that of CH by the factor $\liv/\lambda$, a correction that arises from the vortex gas nature of the flow field.

%\begin{figure}
%%\vspace{-3 cm}
%\includegraphics[width=9 cm]{figz/ell_vs_drag_May19.eps} 
%\includegraphics[width=9 cm]{figz/D_vs_drag_May19.eps} 
%%\centerline{\includegraphics[width=7 cm]{figz/figscaling.eps} }
%\caption{Dimensionless mixing length $\ell_*$ and diffusivity $D_*$ as functions of dimensionless drag, for both linear and quadratic drag. Symbols correspond to numerical simulations, while the solid lines are the predictions [\ref{ellthlinear}], [\ref{Dthlinear}], [\ref{scalingellquad}] and [\ref{scalingDquad}] from the vortex gas scaling theory. \label{figscaling}}
%\end{figure}

\begin{figure*}[t!]
\centerline{\includegraphics[width=9 cm]{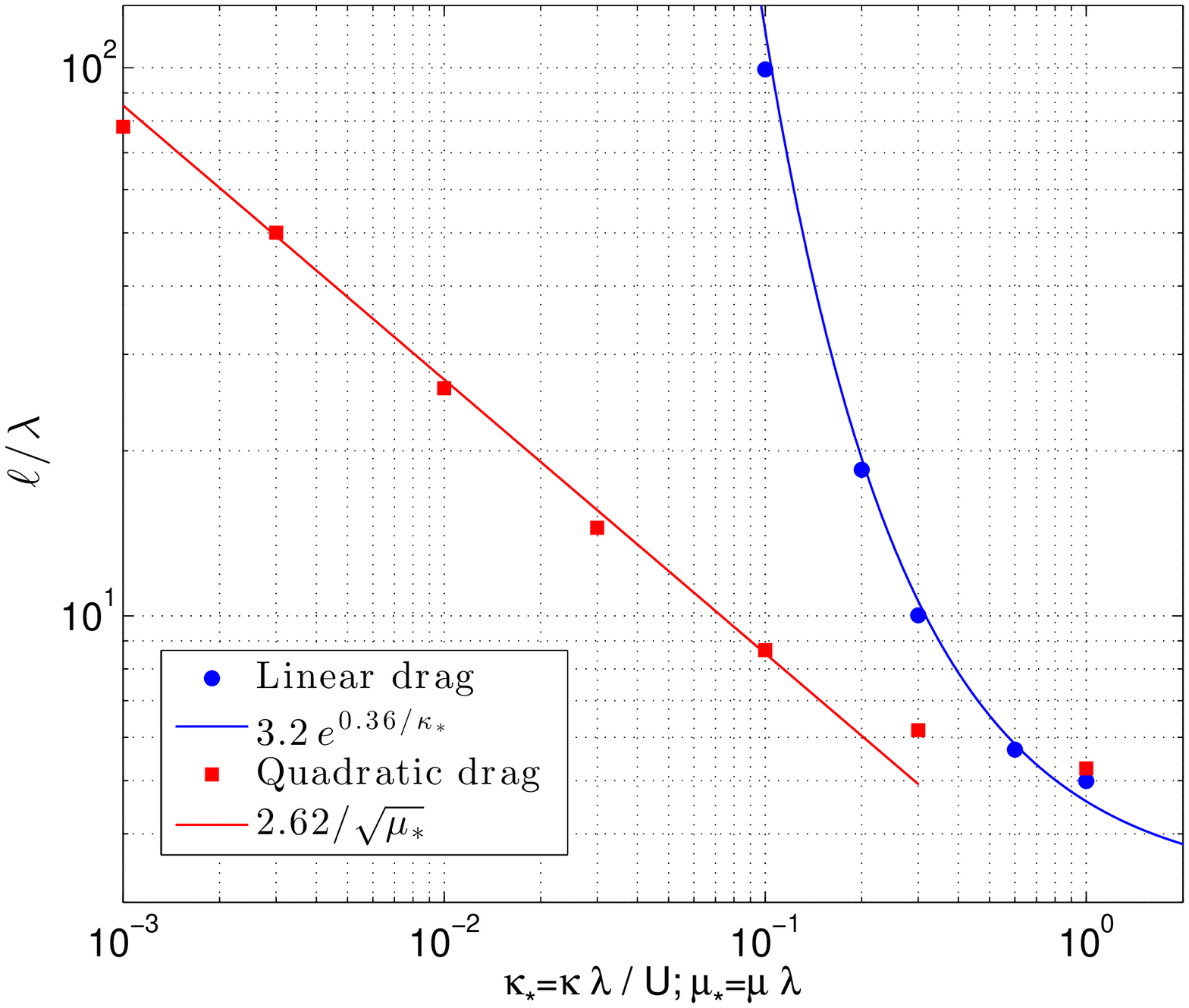} 
\includegraphics[width=9 cm]{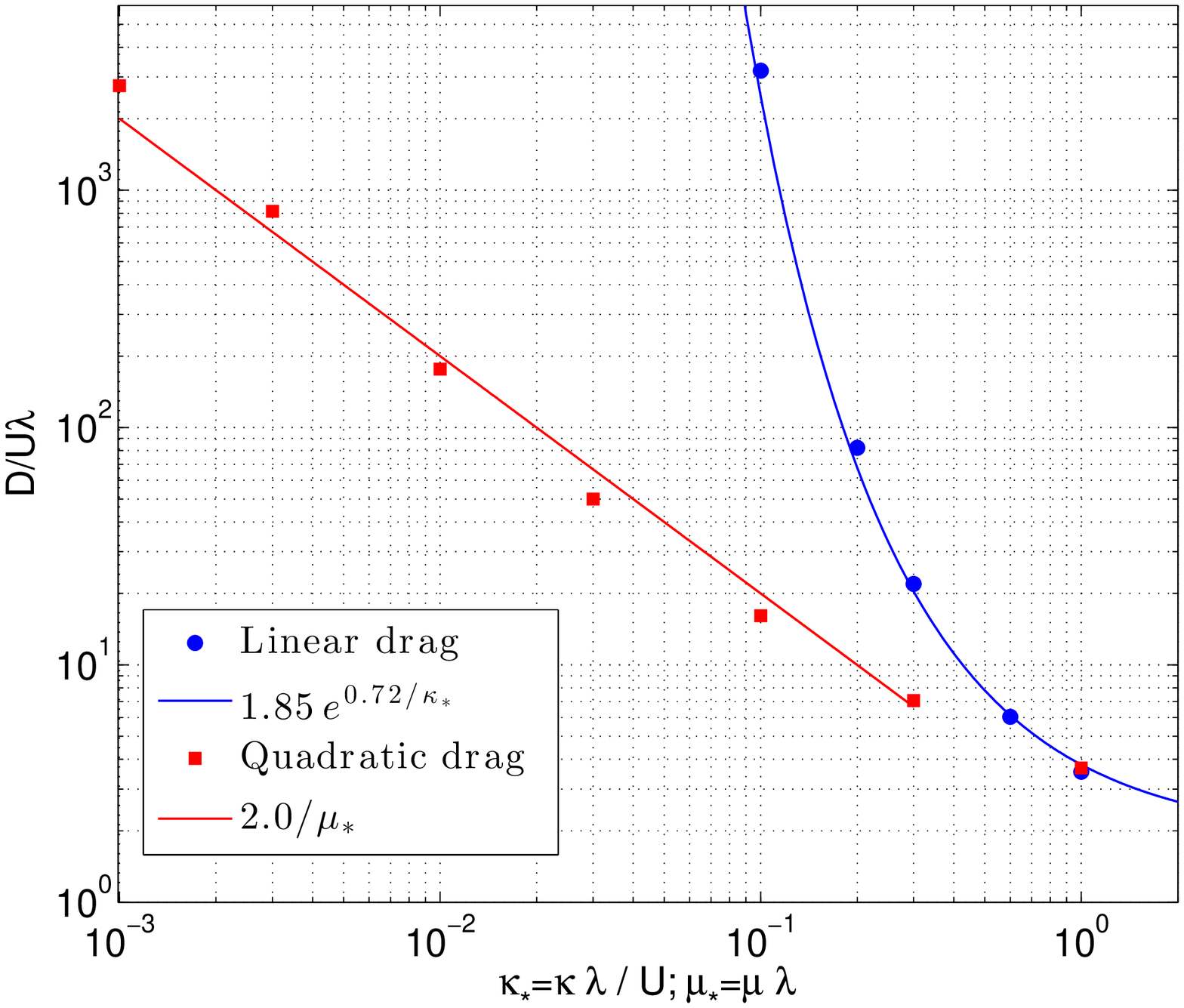} }
%\centerline{\includegraphics[width=7 cm]{figz/figscaling.eps} }
\vspace{2 cm}
\caption{Dimensionless mixing length $\ell_*$ and diffusivity $D_*$ as functions of dimensionless drag, for both linear and quadratic drag. Symbols correspond to numerical simulations, while the solid lines are the predictions [\ref{ellthlinear}], [\ref{Dthlinear}], [\ref{scalingellquad}] and [\ref{scalingDquad}] from the vortex gas scaling theory. \label{figscaling}}
\end{figure*}

The next steps of the scaling theory are common to linear and quadratic drag. As in any mixing-length theory, we will express the diffusion coefficient $D$ as the product of the mixing-length and a typical velocity scale. In the vortex gas regime, one can anticipate that the mixing-length $\ell$ scales as the typical inter-vortex distance $\liv$, an intuition that will be confirmed by equation [\ref{ellvselliv}] below. However, a final relationship for the relevant velocity scale is more difficult to anticipate, as we have seen that the various barotropic velocity moments scale differently. The goal is thus to determine this velocity scale through a precise description of the transport properties of the assembly of vortices.

Stirring of a tracer like temperature takes place at scales larger than the stirring rods, in our problem the vortices of size $\lambda$. \cor{At scales much larger than $\lambda$}, the $\tau$ equation [\ref{eqtau}] reduces to~\cite{Salmon80,Larichev,Thompson}:
\begin{equation}
\partial_t \tau + J(\psi, \tau)= U \psi_x -\nu \Delta^4 \tau \, . \label{tauLS}
\end{equation}
$U \psi_x$ represents the generation of $\tau$--fluctuations through stirring of the large-scale temperature gradient $-U$, and the Jacobian term represents the advection of $\tau$--fluctuations by the barotropic flow. Equation [\ref{tauLS}] is thus that of a passive scalar with an externally imposed uniform gradient $-U$ stirred by the barotropic flow. %We focus on equation [\ref{tauLS}] because it is simpler than the full $\tau$ equation, but also because it shows that the resulting scaling relations hold for the mixing-length and diffusivity of passive tracers as well. 
\cor{To check the validity of this analogy, we have implemented such passive-tracer dynamics into our numerical simulations: in addition to solving equations [\ref{eqpsi}] and [\ref{eqtau}], we solve equation [\ref{tauLS}] with $\tau$ replaced by the concentration $c$ of a passive scalar, and $-U$ replaced by an imposed meridional gradient $-G_c$ of scalar concentration. In the low-drag simulations, the resulting passive scalar diffusivity $D_c=\la \psi_x c \ra/G_c$ equals the temperature diffusivity $D$  within a few percents, whereas $D$ is significantly lower than $D_c$ for larger drag, when the inter-vortex distance becomes comparable to $\lambda$. This validates our assumption that the diffusivity is mostly due to flow structures larger than $\lambda$ in the low-drag regime, whose impact on the temperature field is accurately captured by the approximate equation [\ref{tauLS}]. We can thus safely build intuition into the behavior of the temperature field by studying equation [\ref{tauLS}].}

%As a matter of fact, we implemented passive-tracer dynamics into our low-drag numerical simulations and found that the diffusivity and mixing-length of a passive tracer advected by the barotropic flow are equal to those of the temperature field (within the numerical error bars). We can thus safely build intuition into the behavior of the latter by studying equation [\ref{tauLS}].

A natural first step would be to compute the heat flux associated with a single steady vortex. However, this situation turns out to be rather trivial: the vortex stirs the temperature field along closed circles until it settles in a steady state that has a vanishing projection onto the source term $U \psi_x$, and the resulting heat flux $\la \psi_x \tau \ra$ vanishes up to hyperviscous corrections. Instead of a single steady vortex, the simplest heat carrying configuration is a vortex dipole, such as the one sketched in Fig.~\ref{figdipole}a: two vortices of opposite circulations $\pm \Gamma$ are separated by a distance $\ell_{iv}$ much larger than their core radius $r_\text{core} \sim \lambda$. This dipole mimics the two nearest vortices of any given fluid element, which we argue is a sufficient model to capture the qualitative transport properties of the entire vortex gas. Without loss of generality, the vortices are initially aligned along the zonal axis, and, as a result of their mutual interaction, they travel in the $y$ direction at constant velocity $\Gamma/2\pi \ell_{iv}$. For the configuration sketched in Fig.~\ref{figdipole}a the meridional velocity is positive between the two vortices and becomes negative at both ends of the dipole. For positive $U$, this corresponds to a heat source between the vortices, and two heat sinks away from the dipole. These heat sources and sinks are positively correlated with the local meridional barotropic velocity, so that there is a net meridional heat flux $\la \psi_x \tau \ra$ associated with this configuration. We have integrated numerically equation [\ref{tauLS}] for this moving dipole, over a time $\ell_{iv}/V$, which corresponds to the time needed for the dipole to travel a distance $\ell_{iv}$. This is the typical distance travelled by these two vortices before pairing up with other vortices inside the gas. Panels \ref{figdipole}c,d show the resulting temperature field and local flux $\psi_x \tau$ at the end of the numerical integration (see the SI for details). %\cor{An immediate observation is that the temperature and local heat flux take significant values outside the vortex cores. This may come as a surprise to the reader familiar with TY, who argue that the heat transport is due to the vortices carrying their core temperature anomaly as they drift in the meridional direction. Because the vortex cores cover a negligible fraction of the fluid domain, we argue instead that heat transport takes place predominantly in the inter-vortex region of the dilute vortex gas.}
A suite of numerical simulations for such dipole configurations indicates that, at the end time of the numerical integration, the local mixing length and diffusivity obey the scaling relations:
\begin{eqnarray}
\ell & \sim & \ell_{iv} \, , \label{ellvselliv} \\
%D & \sim & \ell_{iv} V \sim \Gamma \label{Dtemp}\, ,
D & \sim & \ell_{iv} V \label{Dtemp}\, ,
\end{eqnarray}
while the variance and third-order moment of the vortex dipole flow field satisfy [\ref{velvariance}] and [\ref{vel3}]  at every time. It is interesting that the velocity scale arising in the diffusivity [\ref{Dtemp}] is $V$ and not the rms velocity $\la {\bf u}^2 \ra^{1/2}$. This is because the fluid elements that are trapped in the immediate vicinity of the vortex cores do not carry heat, in a similar fashion that a single vortex is unable to transport heat. Only the fluid elements located at a fraction of $\ell_{iv}$ away from the vortex centers carry heat, and these fluid elements have a typical velocity $V$.

The relations [\ref{ellvselliv}] and [\ref{Dtemp}] hold for any passive tracer. However, temperature is an active tracer, so that the velocity scale in-turn depends on the temperature fluctuations, providing the fourth scaling relation. This relation can be derived through a simple heuristic argument: consider a fluid particle, initially at rest, that accelerates in the meridional direction by transforming potential energy into barotropic kinetic energy by flattening the density interface as a result of baroclinic instability. In line with the standard assumptions of a mixing-length model, we assume that the fluid particle travels in the meridional direction over a distance $\ell$, before interacting with the other fluid particles. Balancing the kinetic energy gained over the distance $\ell$ \cor{with the difference in potential energy between two fluid columns  a distance $\ell$ apart}, we obtain the final barotropic velocity of the fluid element: $v_f \sim U \ell / \lambda$. This velocity estimate does not hold for the particles that rapidly loop around a vortex center, with little changes in APE; it holds only for the fluid elements that travel in the meridional direction, following a somewhat straight trajectory (these fluid elements happen to be the ones that carry heat, according to the dipole model described above). Such fluid elements have a typical velocity $V$, which we identify with $v_f$ to obtain: 
\begin{eqnarray}
V \sim U \ell / \lambda \, . \label{ff}
\end{eqnarray}
\cor{A similar relation was derived by Green \cite{Green}, who computes the kinetic energy gained by flattening the density interface over the whole domain. In the present periodic setup the mean slope of the interface is imposed, and the estimate [\ref{ff}] holds locally for the heat-carrying fluid elements travelling a distance $\ell$ instead. The estimate [\ref{ff}] is also} reminiscent of the ``free-fall'' velocity estimate of standard upright convection, where the velocity scale is estimated as the velocity acquired during a free-fall over one mixing-length
%, i.e, $\sqrt{\alpha g \Delta T \ell} \sim \sqrt{\alpha g G \ell^2}$, where $\alpha$ is the thermal expansion coefficient, $g$ is gravity, $\Delta T$ is the temperature difference over one mixing length and $G$ is the background temperature gradient
~\cite{Spiegel63,Spiegel,Gibert}. The conclusion is that the typical velocity is directly proportional to the mixing length. The baroclinic instability is sometimes referred to as slant-wise convection, and the velocity estimate [\ref{ff}] is the corresponding ``slant-wise free fall'' velocity.  To validate [\ref{ff}], one can notice that, when combined with [\ref{ellvselliv}] and [\ref{Dtemp}], it leads to the simple relation:
\begin{eqnarray}
D_* \sim \ell_*^2  \, . \label{Dell2}
\end{eqnarray}
Anticipating the numerical results presented in Fig.~\ref{figscaling}, this relation is well satisfied in the dilute low-drag regime, $\ell \gtrsim 10 \lambda$, the solid lines in both panels being precisely related by [\ref{Dell2}] above. %A closer inspection indicates that the ratio of $D_*/\ell_*^2$ varies by at most a factor of two over the low-drag numerical data, while both sides of equation [\ref{Dell2}] vary by two orders of magnitude. 
A relation very close to [\ref{Dell2}] was reported by Larichev \& Held using turbulent-cascade arguments \cite{Larichev}. Their relation is written in terms of an ``energy containing wavenumber'' instead of a mixing-length. If this energy containing wavenumber is interpreted to be the inverse inter-vortex distance of the vortex-gas model, then their relation becomes identical to [\ref{Dell2}]. 
%However, following TY a better estimate of the energy containing wavenumber may be the wavenumber at which the barotropic energy spectrum peaks. With that definition, a close inspection indicates that the relation of Larichev \& Held differs from [\ref{Dell2}] by a logarithmic term in $\ell_*$. While this term can be safely neglected in a first approximation in the case of quadratic drag, it turns out to play a crucial role in the case of linear drag, and the precise form [\ref{Dell2}] need be retained.

\begin{figure*}[h!]
    \centerline{\includegraphics[width=18 cm]{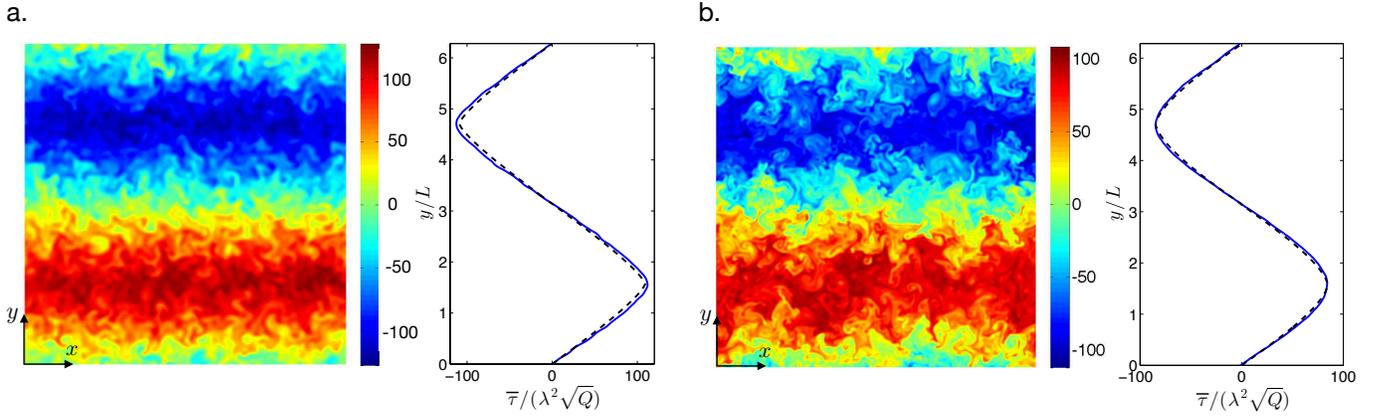} }
   \caption{\textbf{Testing the diffusive closure.} Snapshots and meridional profiles of the dimensionless temperature $\tau/\lambda^2 Q^{1/2}$. The solid lines are the zonal and time mean from the numerical simulations, while the dashed lines are the theoretical expressions [\ref{taulin}] and [\ref{tauquad}]. a. Linear drag, with $\kappa/Q^{1/2}=0.5$ and $\lambda/L=0.02$. b. Quadratic drag, with $\mu_*=10^{-2}$ and $\lambda/L=0.01$.    \label{figheat}}
\end{figure*}

%The last key relation of this scaling theory is the energy budget: the meridional heat flux corresponds to a rate of release of APE, which is balanced by frictional dissipation of kinetic energy in statistically steady state. The contribution from the barotropic flow dominates this frictional dissipation in the low-drag asymptotic limit. For linear drag, the energy budget thus reads (see e.g. TY, and the SI):
%\begin{eqnarray}
%\kappa \la {\bf u}^2 \ra \simeq \frac{U \la \psi_x \tau \ra}{\lambda^2} = \frac{D U^2}{\lambda^2} \, .
%\end{eqnarray}

The four relations needed to establish the scaling theory are [\ref{Ebudget}], [\ref{ellvselliv}], [\ref{Dtemp}] and [\ref{ff}]. In the case of linear drag, their combination leads to $\log(\ell_*) \sim 1/\kappa_*$, or simply:
\begin{equation}
\ell_* = c_1 \, \exp\left({\frac{c_2}{\kappa_*}}\right) \, , \label{ellthlinear}
\end{equation} 
where $c_1$ and $c_2$ are dimensionless constants. The vortex gas approach thus provides a clear theoretical explanation to the exponential dependence of $\ell$ on inverse drag reported by TY, which is shown to stem from the logarithmic factor in [\ref{velvariance}] for the dissipation of kinetic energy. It is remarkable that these authors could extract the correct functional dependence of $\ell_*$ with $\kappa_*$ from their numerical simulations. We have performed similar numerical simulations, in large enough domains to avoid finite-size effects, and at low enough hyperviscosity to neglect hyperdissipation in the kinetic energy budget. The numerical implementation of the equations, as well as the parameter values of the various numerical runs, are provided in the SI. In Fig.~\ref{figscaling}, we plot $\ell_*$ as a function of $\kappa_*$. We obtain an excellent agreement between the asymptotic prediction [\ref{ellthlinear}] and our numerical data using $c_1=3.2$ and $c_2=0.36$. The dimensionless diffusivity is deduced from $\ell_*$ using the relation [\ref{Dell2}], which leads to:
\begin{equation}
D_* = c_3 \, \exp\left( {\frac{2  c_2}{\kappa_*}}\right) \, . \label{Dthlinear}
\end{equation} 
Once again, upon choosing $c_3=1.85$ this expression is in excellent agreement with the numerical data, see Fig.~\ref{figscaling}.

When linear friction is replaced by quadratic drag, only the energy budget [\ref{Ebudget}] is modified. As can be seen in equation [\ref{vel3}], the main difference is that quadratic drag operates predominantly in the vicinity of the vortex cores, which has a direct impact on the scaling behaviors of $\ell_*$ and $D_*$. Indeed, combining [\ref{Ebudget}], [\ref{ellvselliv}], [\ref{Dtemp}] and [\ref{ff}] yields:
\begin{eqnarray}
\ell_* = \frac{c_4}{\sqrt{\mu_*}} \, , \label{scalingellquad}
\end{eqnarray}
which, using [\ref{Dell2}], leads to the diffusivity:
\begin{eqnarray}
D_* = \frac{c_5}{\mu_*} \, . \label{scalingDquad}
\end{eqnarray}
Using the values $c_4=2.62$ and $c_5=2.0$, the predictions are again in very good agreement with the numerical data, although the convergence to the asymptotic prediction for $D_*$ seems somewhat slower for this configuration, see Fig.~\ref{figscaling}.

\section*{Using these scaling-laws as a local closure}

We now wish to demonstrate the skill of these scaling-laws as local diffusive closures in situations where the heat-flux and the temperature gradient have some meridional variations. For simplicity, we consider an imposed heat flux with a sinusoidal dependence in the meridional direction $y$. The \cor{modified} governing equations for the potential vorticities $q_{1;2}$ of each layer are:
\begin{eqnarray}
\partial_t q_1 + J(\psi_1,q_1) & = &  Q \sin(y/L) -\nu \Delta^4 q_1 \, , \label{q1flux}\\
\partial_t q_2 + J(\psi_2,q_2) & = & -Q \sin(y/L) -\nu \Delta^4 q_2 + \text{drag} \label{q2flux} \, .
\end{eqnarray}
It becomes apparent that the $Q$-terms represent a heat flux when the governing equations are written for the (total) baroclinic and barotropic streamfunctions $\tau$ and $\psi$: the $\tau$-equation, obtained by subtracting [\ref{q2flux}] from [\ref{q1flux}] and dividing by two, has a source term $Q \sin(y/L)$ that forces some meridional temperature structure. By contrast, the $\psi$-equation obtained by adding [\ref{q1flux}] and [\ref{q2flux}] has no source terms. The goal is to determine the temperature profile associated with the imposed meridionally dependent heat flux. This slantwise convection forced by sources and sinks is somewhat similar to standard upright convection forced by sources and sinks of heat \cite{Lepot,Bouillaut}. %While in both cases the link between the heat flux and the temperature gradients does not involve the small-scale molecular diffusivities, we have seen that for slantwise convection it strongly involves the large-scale drag. 
We focus on the  statistically steady state by considering a zonal and time average, denoted as $\overline{\cdot}$. Neglecting the dissipative terms, the average of both equations [\ref{q1flux}] and [\ref{q2flux}] leads to:
\begin{eqnarray}
Q \sin(y/L)=-\frac{1}{\lambda^2}\partial_y \overline{\psi_x \tau} \, . \label{fluxbalance}
\end{eqnarray}
Provided the imposed heat flux varies on a scale $L$ much larger than the local mixing-length $\ell$, we can relate the local flux $\overline{\psi_x \tau}(y)$ to the local temperature gradient $U(y)=\partial_y{\overline{\tau}}$ by the diffusive relation $\overline{\psi_x \tau}(y)=D U(y)= D_* \lambda |U(y)| U(y)$. In the case of quadratic drag, inserting this relation into [\ref{fluxbalance}] and substituting the scaling-law [\ref{scalingDquad}] for $D_*(\mu_*)$ yields:
\begin{eqnarray}
-\frac{c_5}{\mu_*} \partial_y \left[  |\partial_y \overline{\tau}| \partial_y \overline{\tau} \right]  =  Q \sin(y/L) \, .
\end{eqnarray}
In terms of the dimensionless temperature $\tau_*=\tau/\lambda^2 \sqrt{Q}$, the solution to this equation is:
\begin{eqnarray}
%\overline{\tau}_*(y/L) = \left( \frac{L }{\lambda} \right)^{3/2} \frac{2 \sqrt{\mu_*}}{\sqrt{c_5}} {\cal E} \left( \frac{y}{2L} | 2 \right) \, , 
\overline{\tau}_*(y/L) = 2 \left( \frac{L }{\lambda} \right)^{3/2} \sqrt{\frac{{\mu_*}}{{c_5}}} \, {\cal E} \left( \frac{y}{2L} | 2 \right) \, , \label{tauquad}
\end{eqnarray}
where ${\cal E}$ denotes the incomplete elliptic integral of the second kind. Expression [\ref{tauquad}] holds for $y/L \in [-\pi/2;\pi/2]$, the entire graph being easily deduced from the fact that $\overline{\tau}_*(y/L)$ is symmetric to a translation by $\pi$ accompanied by a sign change.

In the case of linear drag, we substitute the scaling-law  [\ref{Dthlinear}] for $D_*(\kappa_*)=D_*(\lambda \kappa/|\partial_y \overline{\tau}| )$ instead. The integration of the resulting ODE yields the dimensionless temperature profile:
\begin{eqnarray}
%\overline{\tau}_*(y/L) = \frac{\kappa L}{c_2 \lambda \sqrt{Q}} \int_0^{y/L} {\cal W} \left( \frac{c_2}{\sqrt{c_3}} \sqrt{\frac{L}{\lambda}} \frac{\sqrt{Q}}{\kappa} \sqrt{\cos s} \right) \mathrm{d}s \, ,\label{taulin}
\overline{\tau}_*(y/L) = \frac{\kappa L}{c_2 \lambda \sqrt{Q}} \int_0^{y/L} {\cal W} \left( \frac{c_2}{\kappa} \sqrt{\frac{ L Q}{c_3 \lambda } \, \cos s  } \,  \right) \mathrm{d}s \, ,\label{taulin}
\end{eqnarray}
where ${\cal W}$ denotes the Lambert function. Once again, [\ref{taulin}] holds for $y/L \in [-\pi/2;\pi/2]$,  the entire graph being easily deduced from the fact that $\overline{\tau}_*(y/L)$ is symmetric to a translation by $\pi$ accompanied by a sign change.

To test these theoretical predictions, we solved numerically equations [\ref{q1flux}-\ref{q2flux}] inside a domain $(x,y)\in[0; 2\pi L]^2$ with periodic boundary conditions, for both linear and quadratic drag. We compute the time and zonally averaged temperature profiles and compare them to the theoretical predictions, \cor{using the values of the parameters $c_{1;2;3;4;5}$ deduced above.} In Fig.~\ref{figheat}, we show snapshots of the temperature field in statistically steady state, together with meridional temperature profiles. The predictions [\ref{tauquad}] and [\ref{taulin}] are in excellent agreement with the numerical results for both linear and quadratic drag, and this good agreement holds provided the various length scales of the problem are ordered in the following fashion: $\lambda \ll \ell \ll L$. The first inequality corresponds to the dilute vortex gas regime for which the scaling theory is established, while the second inequality is the scale separation required for any diffusive closure to hold. For fixed $L/\lambda$, the first inequality breaks down at large friction, $\kappa_* \sim 1$ or $\mu_* \sim 1$, where the system becomes a closely packed ``vortex liquid''~\cite{Arbic2004,Arbic}. The second inequality breaks down at low friction, when $\ell \sim L$. From the scaling-laws [\ref{ellthlinear}] and [\ref{scalingellquad}], this loss of scale separation occurs for $\kappa_* \lesssim 1/\log(L/\lambda)$ and $\mu_* \lesssim (\lambda/L)^2$, respectively for linear and quadratic drag.

\section*{Discussion}

The vortex gas description of baroclinic turbulence allowed us to derive predictive scaling-laws for the dependence of the mixing-length and diffusivity on bottom friction, and to capture the key differences between linear and quadratic drag. \cor{The scaling behavior of the diffusivity of baroclinic turbulence appears more ``universal'' than that of its purely barotropic counterpart. This is likely because many different mechanisms are used in the literature to drive purely barotropic turbulence. For instance, the power input by a steady sinusoidal forcing \cite{TsangYoung,Tsang} strongly differs from that input by forcing with a finite \cite{Maltrud} or vanishing \cite{Grianik} correlation time, with important consequences for the large-scale properties and diffusivity of the resulting flow. By contrast, baroclinic turbulence comes with its own injection mechanism -- baroclinic instability -- and the resulting scaling-laws depend only on the form of the drag.}
We demonstrated the skills of these scaling-laws when used as local parameterizations of the turbulent heat transport, in situations where the large-scale forcing is inhomogeneous. 
While this theory provides some qualitative understanding of turbulent heat transport in planetary atmospheres, it should be recognized that the scale separation is at best moderate in Earth atmosphere, where meridional changes in the Coriolis parameter also drive intense jets. On the other hand, our firmly footed scaling theory could be the starting point towards a complete parameterization of baroclinic turbulence in the ocean, a much-needed ingredient of global ocean models. Along the path, one would need to adapt the present approach to models with multiple layers, possibly going all the way to a geostrophic model with continuous density stratification, or even back to the primitive equations. The question would then be whether the vortex gas provides a good description of the equilibrated state in these more general settings. Even more challenging would be the need to include additional physical ingredients in the scaling theory: the meridional changes in $f$ mentioned above, but also variations in bottom topography, and surface wind stress. Whether the vortex gas approach holds in those cases will be the topic of future studies.

\paragraph{Data availability.} The data associated with this study are available within the paper and SI.

\showmatmethods{} % Display the Materials and Methods section

\acknow{Our work is supported by the generosity of Eric and Wendy Schmidt by recommendation of the Schmidt Futures program, and by the National Science Foundation under grant AGS-6939393. This research is also supported by the European Research Council (ERC) under grant agreement FLAVE 757239.}

\showacknow{} % Display the acknowledgments section

% Bibliography
\bibliography{pnas-sample}

\end{document}